\documentclass[%
aps,pre,twocolumn,superscriptaddress,groupedaddress
]{revtex4}

\usepackage{graphicx}
\usepackage{dcolumn}
\usepackage{bm}
\usepackage{amsmath}
\usepackage{amssymb}
\usepackage{braket}
\usepackage[utf8]{inputenc}
\usepackage[english]{babel}

\begin{document}

\preprint{APS/123-QED}

\title{Complex Solitary Waves and Soliton Trains in KdV and mKdV Equations }

\author{Subhrajit Modak}
\email{modoksuvrojit@gmail.com}
\affiliation{Indian Institute of Science Education and Research Kolkata, Mohanpur 741246, India}%
\author{Akhil P. Singh}
\affiliation{BITS Pilani, K.K. Birla Goa Campus, Goa 403726, India }
\author{P. K. Panigrahi}
\affiliation{Indian Institute of Science Education and Research Kolkata, Mohanpur 741246, India}%

\date{\today}

\begin{abstract}
We demonstrate the existence of complex solitary wave and periodic solutions of the Kortweg de-vries (KdV) and modified Kortweg de-Vries (mKdV) equations. The solutions of the KdV (mKdV) equation appear in complex-conjugate pairs and are even (odd) under the simultaneous actions of parity ($\cal{P}$) and time-reversal ($\cal{T}$) operations. The corresponding localized solitons are  hydrodynamic analogs of Bloch soliton in magnetic system, with asymptotically vanishing intensity. The $\cal{PT}$-odd complex soliton solution is shown to be iso-spectrally connected to the fundamental $sech^2$ solution through supersymmetry.


\end{abstract}

\maketitle


\section{\label{sec:level1}Introduction}

The celebrated Kortweg-de Vries (KdV) equation is a well-studied non-linear dynamical system, first evoked for the description of 
\emph{solitary waves} \cite{1} in shallow water \cite{2}. It has,  since then, found much applications \cite{3}. An integrable 
model \cite{4}, it arises from the compatibility condition of two linear equations, the well-known Lax-pair \cite{5}. The fact that one 
of them is the linear Schr\"odinger operator, connects this non-linear equation to quantum mechanical eigenvalue problem of 
the reflectionless potential \cite{6,7}. It has both localized and periodic cnoidal wave solutions, which appear in the Lax equation 
as potentials, giving rise to  bound states \cite{8} and band structure \cite{9}, respectively. A number of methods, viz., inverse 
scattering \cite{10}, Hirota bilinear \cite{11} etc., have been developed to find the general multi-soliton solutions of the KdV 
hierarchy. The KdV solutions are connected by the Miura transformation \cite{12} to the modified KdV (mKdV) equation \cite{13}, 
which also has found diverse physical applications. It appears in the description of
van Alfv\'en waves in collisionless plasma \cite{14}, phonons in anharmonic lattice \cite{15}, interfacial waves in two-layer liquid with gradually varying depth \cite{16}, transmission lines in Schottky barrier \cite{17} and ion  acoustic  solitons \cite{18,19,20}, to mention a few. 
\paragraph*{}In addition to the fundamental localized soliton solution: $-2\alpha^2 sech^2[\alpha(x-4\alpha^2 t)]$ and its periodic counterpart, $-2m\alpha^2 cn^2[\alpha(x-4(2m-1)\alpha^2 t)]$, recently real superposed solutions have been found to satisfy the KdV equation \cite{22,23}.
Here, we show that KdV equation possesses $\cal{PT}$-symmetric, localized and cnoidal wave solutions, which appear in complex conjugate pairs. 
Interestingly, the sum of the paired solution is also a solution of the KdV equation, whereas the difference is not.
The existence of these solutions can be traced to the fact that, there are two distinct mKdV equations,
\begin{eqnarray}
v_{1,t}-6v_{1}^2v_{1,x}+v_{1,xxx}&=&0\nonumber\\ 
\text{and}~~~v_{2,t}+6v_{2}^2v_{2,x}+v_{2,xxx}&=&0,\nonumber
\end{eqnarray}
whose solutions are related through $v\rightarrow iv$. The Miura transformation for the solution of KdV, $u=v^2\pm v_{x}$, then implies that, $u=-v^2\pm i v_{x}$, is also a solution. We find complex $\cal{PT}$-odd solutions for the mKdV equation, where the sum and the differences of the pair are also solutions. This, in turn, generates more general complex superposed solutions for the KdV equation. The complex soliton  solutions  are analogs of Bloch solitons in magnetic systems \cite{24}. The $\cal{PT}$-symmetric soliton solution is shown to be isospectrally related to the fundamental solution $sech^2x$, in the Lax equation.
\section{\label{sec:level1}Complex Superposed solutions}

We start here with the KdV equation,
\begin{equation}
u_t -6 u u_x + u_{xxx}=0,\nonumber
\end{equation}
where, $u=u(x,t)$, $u_t= \frac{\partial u }{\partial t}$, $u_x= \frac{\partial u}{\partial x}$ and $u_{xxx}= \frac{\partial^3 u}{\partial x^3}$.

\subsection{\label{sec:level2}$cn^2 \pm i sn dn$ type solution}
It is straightforward to check that, the following pair of complex periodic solutions satisfy the KdV equation:

\begin{center}
$u(x,t)=A cn^2 (\zeta,m) + i B sn (\zeta,m) dn(\zeta,m)$,
\end{center}

where $\zeta=\alpha(x-c\alpha^2 t)$, provided $A=-m \alpha^2$, $B=\pm \sqrt{m} \alpha^2$ and $c= (2m-1)$, $m$ being the modulus parameter. Evidently, velocity c exhibits two disjoint domains: for $\frac{1}{2} < m \leq 1$, the solution is right moving, while for the remaining half, $0 < m < \frac{1}{2}$, it is left moving. It is interesting to note that, the sum of the complex pair is the well-known cnoidal wave solution, with $c=4(2m-1)$ and $A=-2m\alpha^2$, whereas the difference of the pair is no longer a solution. The real ($u_{R}$) and imaginary parts ($u_{I}$) 
of the solutions are plotted in Fig. 1, which are even and odd functions of the argument, respectively. The fundamental soliton has higher intensity in comparison to the superposed solution.

%

\begin{center}
\includegraphics[width=9cm]{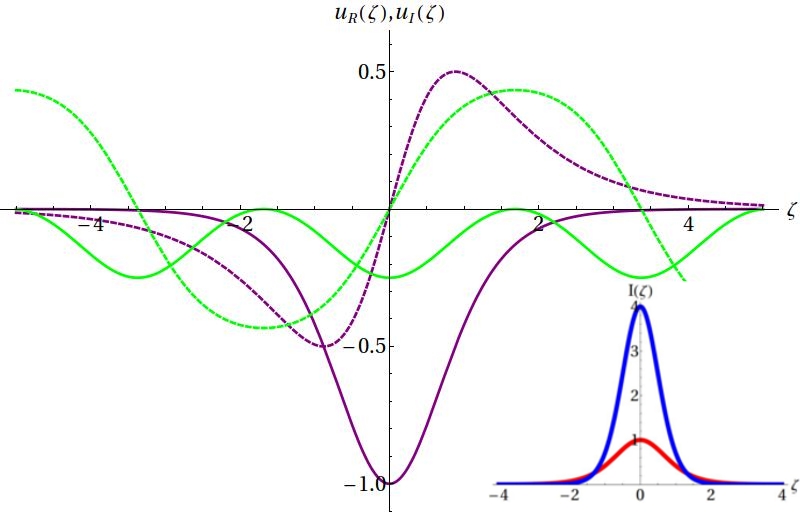}\\
\end{center}
FIG. 1: Plots of real (solid) and imaginary (dashed) parts of $u(\zeta)=cn^2(\zeta)+i sn(\zeta)dn(\zeta)$, for $m=1$ (purple)
and $m=0.25$ (green). The inset shows intensities
of superposed (red) and fundamental (blue) soliton solutions, with the latter being larger. For all the cases, $\alpha=1$.

\subsection{\label{sec:level2}$cn^2 \pm i sn cn$ type solutions}
The following factorizable, superposed solution also satisfies KdV equation:
\begin{center}
$u(x,t)=A cn^2 (\zeta,m) + i B sn (\zeta,m) cn(\zeta,m) + \beta \alpha^2$,
\end{center}
provided $ A = -m \alpha^2$, $B= \pm A$ and $c=(5m-4)-6\beta$. As in the previous case, here also, oppositely propagating modes occupy two different domains of $m$. The real and imaginary parts of the solutions are plotted in Fig. 2. Here also, addition of the paired localized solutions ($m=1$) yields the fundamental soliton with higher intensity.\newline

%

\begin{center}
\includegraphics[width=9cm]{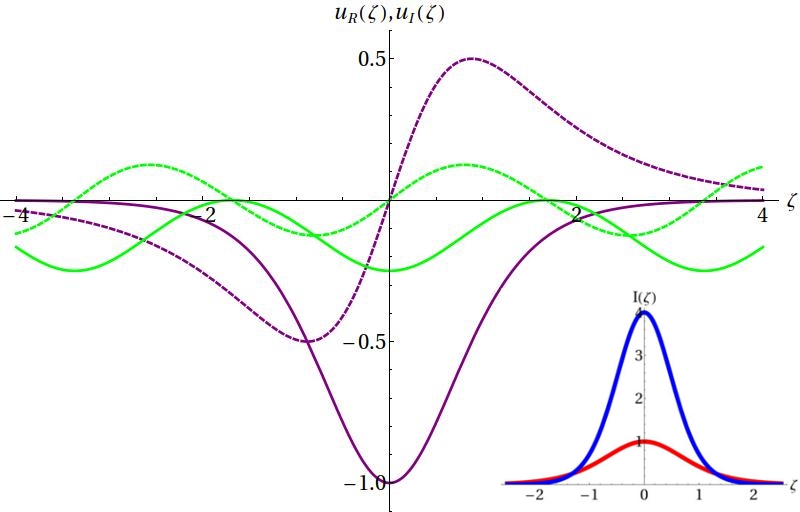}\\
\end{center}
FIG. 2: Plots of real (solid) and imaginary (dashed) parts of $u(\zeta)=cn^2(\zeta)+i sn(\zeta)cn(\zeta)$, for $m=1$ (purple)
and $m=0.25$ (green). The real parts are symmetric, whereas the imaginary ones are antisymmetric about $\zeta$. For $\alpha=1$, the inset shows intensities
of superposed (red) and fundamental (blue) soliton solutions, with the latter being larger. 
\subsection{\label{sec:level2}Remarks and observations}
For $m$ = 1, superposed pairs reduce to $\cal{PT}$-symmetric complex form:  $u(x; t) = \text{sech}^2 \zeta \pm i \text{sech} \zeta \text{tanh} \zeta$. Use of the Cole-Hopf transformation: $v=\frac{\Psi_x}{\psi}$, in one of the Miura route: $u=v^2+v_{x}$, 
 yields $u=\frac{\psi_{xx}}{\psi}$. The Galilean invariance of the KdV equation \cite{28}  allows a constant shift in $u$, leading to
\begin{center}
$-\psi_{xx}+[\lambda+u(x,t)]\psi=0$,
\end{center}
 the one dimensional Schr\"odinger equation with a Scarf-type $\cal{PT}$-symmetric potential \cite{41,29,26}. Soliton solutions can be generated through iso-spectral deformation of the potential \cite{30a,30}, wherein both the wave function and the potential can change, leaving the spectrum invariant. Depending on the number of bound states, this iso-spectral flow introduces parameters, suitably interpretable as time variables \cite{31}. The complex Scarf potential is known to be iso-spectral to the real $sech^2 x$ potential
\cite{25}, explaining the common real eigenvalues for both these potentials.
\\
\\
KdV also posseses real superposed singular \cite{32} solution: $u(x,t)= \text{A cosech}^2 \zeta+ B \text{cosech} \zeta \text{coth} \zeta$, with $ A = \alpha^2$,  $B = \pm \alpha^2$ and $c = 1$.  

\section{\label{sec:level1}Complex Superposition solutions}
For the mKdV equation,
\begin{center}
$v_t - 6 v^2 v_x + v_{xxx}=0$,
\end{center}
there exist solutions in the form of complex superposition. 
\subsection{\label{sec:level2}$sn \pm i cn$ type solutions}
The following parity odd superposition solution solves mKdV equation:
\begin{center}
$v=A \alpha sn(\zeta,m)+ i B \alpha cn(\zeta,m)$,
\end{center}
provided $A=B=\pm \frac{\sqrt{m}}{2}$ and $c=\frac{m}{2}-1$. Unlike KdV, the sum of the pair,  as well as their difference, simultaneously satisfy the mKdV dynamics. In case of the sum, $v=2A\alpha sn(\zeta,m)$ is an exact solution, when $c=5(m-5)$ and $A=\pm\frac{\sqrt{m}}{2}$, while for the difference, $v=2iB\alpha cn(\zeta,m)$ is also an exact solution, provided $c=(2m-1)$ and $B=\pm\frac{\sqrt{m}}{2}$. These form of solutions generate more general superposed solutions for KdV equation. The real and imaginary parts of the paired solutions are plotted in Fig. 3, which are odd and even functions of the argument, respectively. The corresponding intensity is constant, $I=\frac{m\alpha^2}{4}$.
\begin{center}
\includegraphics[width=9cm]{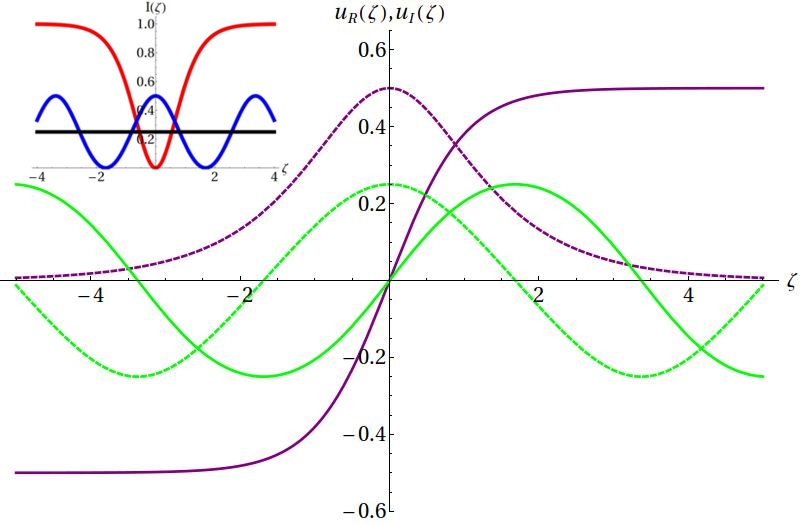} \\
\end{center}
FIG. 3: Plots of real (solid) and imaginary (dashed) parts of $u(\zeta)=sn(\zeta)+icn(\zeta)$, for $m=1$ (purple)
and $m=0.25$ (green). The real parts are antisymmetric, whereas the imaginary ones are symmetric about $\zeta$. The inset shows intensities
of individual (black), added (red) and subtracted (blue) fundamental soliton solutions, with $\alpha=1$.
\subsection{\label{sec:level2}$sn \pm i dn$ type solutions}

Another form of complex periodic pair will satisfy the mKdV equation:

\begin{center}
$v=A \alpha sn(\zeta,m)+ i B\alpha dn(\zeta,m)$,\nonumber
\end{center}
provided $ A=\pm \frac{\sqrt{m}}{2}$, $ B=\pm\frac{1}{2}$ and $c=\frac{1}{2}-m$. In case of $m=1$, one obtains $v(x,t)=\frac{1}{2}(tanh \zeta \pm i sech \zeta)$, which is odd under $\cal{PT}$ operation. It is interesting to note that, the velocity of the pair, $sn\pm i cn$ is $-1/2$ times the velocity of $dn$ solution and the velocity of the pair, $sn\pm i dn$ is $-1/2$ times the velocity of $cn$ solution. Fig. 4 plots the real and imaginary parts of the paired solutions, which again are odd and even functions, respectively. Intensity associated with the excitation is constant, $I=\frac{\alpha^2}{4}$ and is independent of the modulus parameter. Real, singular superposition solutions of the form $v(x,t)=\text{A cosech} \zeta +\text{ B coth} \zeta$, also satisfy mKdV dynamics, with  $B = A=\pm\frac{\alpha}{2}$ and $c = -1/2$.

\begin{center}
\includegraphics[width=9cm]{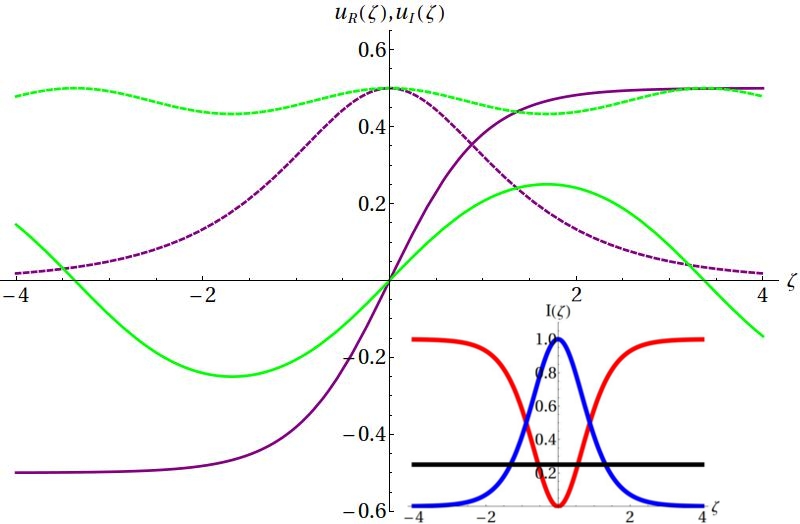} \\
\end{center}
FIG. 4: Plots of real (solid) and imaginary (dashed) parts of $u(\zeta)=sn(\zeta)+idn(\zeta)$, for $m=1$ (purple)
and $m=0.25$ (green). The real parts are antisymmetric, whereas the imaginary ones are symmetric about $\zeta$. The inset shows intensities
of individual (black), added (red) and subtracted (blue) fundamental soliton solutions, with $\alpha=1$.
 \section{Conclusion}
In conclusion, trivially $\cal{PT}$-symmetric KdV equation is shown to posses $\cal{PT}$-symmetric complex solutions,
with asymptotically vanishing intensity for the solitons. In the corresponding Lax equation, the $\cal{PT}$-symmetric potential supports 
real eigenvalues and is iso-spectrally connected to the real potential $sech^2x$, the fundamental soliton solution. Iso-spectral deformation \cite{30} has been useful in generating the multi-soliton solution for the KdV equation. The realization of the same in the case of complex $\cal{PT}$-symmetric potentials needs careful investigation. The fact that in the $\cal{PT}$-symmetric phase , an inner product \cite{77} is defined, may facilitate in obtaining these multi-soliton solutions through iso-spectral deformation. For
the mKdV equation, $\cal{PT}$-odd solutions have been found to be exact solutions, which generate general $\cal{PT}$-symmetric potentials for the KdV equation, through Miura transformation. The KdV and other integrable systems, like Boussinesq hierarchy, manifest in two-dimensional induced gravity \cite{69,49,70} and conformal field theory \cite{42}. The implication of the complex $\cal{PT}$-symmetric solutions in the context of two-dimensional gravity is worth exploring.


\end{document}